\documentclass[%
10pt,
twocolumn,
superscriptaddress,
nofootinbib,
amsmath,
amssymb,
aps,
prd,
]{revtex4-2}

\usepackage{graphicx}
\usepackage{standalone}
\usepackage{float}
\usepackage{dcolumn}
\usepackage{bm}
\makeatletter
\def\tagform@#1{(\textcolor{red}{#1})}
\makeatother
\usepackage[colorlinks=true,linkcolor=red,citecolor=red,urlcolor=red]{hyperref}%

\usepackage{xcolor}
\usepackage{nicefrac}
\usepackage{lipsum}

\usepackage{tikz}
\usepackage{tikz-feynman}
\usepackage{amsmath}
\usepackage{graphicx}
\tikzfeynmanset{compat=1.1.0}
\usetikzlibrary{calc,trees,positioning,arrows,chains,shapes.geometric,decorations.pathreplacing,decorations.pathmorphing,shapes,matrix,shapes.symbols}

\def\beq{\begin{equation}}
\def\eeq{\end{equation}}
\def\bea{\arraycolsep .1em \begin{eqnarray}}
\def\eea{\end{eqnarray}}

\begin{document}

\title{\color{blue}\bf Renormalisation Group approach to General Relativity}

\author{F. Gutiérrez}
\email{fgutierrez@fing.edu.uy}
\affiliation{IFFI, Universidad de la Rep\'ublica, J.H.y Reissig 565, 11300 Montevideo, Uruguay}

\author{K. Falls}
\email{kfalls@fing.edu.uy}
\affiliation{IFFI, Universidad de la Rep\'ublica, J.H.y Reissig 565, 11300 Montevideo, Uruguay}

\author{A. Codello}
\email{alessandro.codello@unive.it}
\affiliation{DMSN, Ca'\ Foscari University of Venice, Via Torino 155, 30172 - Venice, Italy}
\affiliation{IFFI, Universidad de la Rep\'ublica, J.H.y Reissig 565, 11300 Montevideo, Uruguay}

\date{\today}

\begin{abstract}
The detection of gravitational waves has intensified the need for efficient, high-precision modeling of the two-body problem in General Relativity. Current analytical methods, primarily the Post-Minkowskian and Post-Newtonian expansions, are inherently perturbative, while Numerical Relativity remains computationally expensive. In this Letter, we introduce a middle path: an exact renormalization group (RG) equation for classical gravitational systems.
After demonstrating that our equation correctly reproduces the Post-Minkowskian expansion, we show how it easily recovers the 1PN two-body action, bypassing the need for complex three-graviton vertex calculations.
This establishes the exact RG as a powerful new tool for tackling strong-field dynamics in gravity.
%
\end{abstract}

\maketitle

\subsection*{\color{blue}Introduction}\vspace{-0.2cm}

Since the direct observation of gravitational waves by the large interferometers LIGO and VIRGO \cite{Abbott:2016blz,Aasi:2015xik,Acernese:2014mare2} the need to solve strongly coupled problems in general relativity has become increasingly important.
The main effort has focused on finding solutions to the two body problem in the full relativistic theory that describes the inspiralling, merger and ring-down of black hole pairs or other astrophysical relevant situations \cite{AmaroSeoane:2017}.
The statistical matched filtering techniques \cite{Maggiore:2007ulw} -- necessary to distinguish the gravitational wave perturbation from the background noise at the interferometer -- need a rapid solution of the underlying relativistic problem in order to compute the waveforms for the vast parameter space (masses, spin and other properties) that is scanned to match the signal.

In absence of an exact solution, and being the computational route offered by numerical relativity \cite{Baumgarte:2010} quite expensive and slow -- even with present day technology --  analytical efforts have been mostly directed to compute high order corrections within the Post-Minkowskian (PM) and Post-Newtonian (PN) expansions -- which are respectively an expansion around flat space in the gravitational coupling $G_N$ and a low speed expansion in powers of $\frac{v}{c}$.
The field that was born already at the times of Einstein with the first computation of the 1PN action \cite{Einstein:1938yz} and that slowly developed in the second part of the last century \cite{Buonanno:1998gg,Jaranowski:1997ky,Blanchet:2002gz,Poisson:2014book} saw a literal explosion in interest after the aforementioned observations of gravitational waves. Many groups worldwide have applied state-of-the-art techniques from quantum field theory, such as effective field theory ideas (EFT) \cite{Goldberger:2006bd,Porto:2016pyg,Levi:2020pnEFT,Rothstein:2003zf,Gilmore:2008gz,Foffa:2013gsa}; techniques ported from the vast experience accumulated with Feynman integrals \cite{Driesse:2024feo}  and even the most recent amplitudes and scattering methods \cite{Bern:2019crd,Bern:2021tail,BjerrumBohr:2021, Luna:2015} -- all focused to progressively compute higher and higher orders of the PM or PN expansion.

While these calculations have clearly been of crucial importance to the subject they are still  all inherently perturbative. In this Letter we present a middle way between ``exact" numerical relativity and these analytical efforts by presenting an exact renormalization group (RG) equation for classical general relativistic systems. In analogy with statistical mechanics -- where we can resort to Monte Carlo simulation on the numerical side and loop calculation on the perturbative one -- a third option is offered by exact RG equations \cite{Polchinski:1983gv, Wetterich:1989xg,Morris:1993qb}.

After giving an heuristic derivation of the exact RG equation for GR (of which we will a give a more rigorous derivation in the Polchinski RG framework in a companion paper \cite{Paper2}) we start by showing that our RG equation correctly reproduces the PM expansion -- similarly to  quantum RG equations, which are able to correctly recover the loop expansion order by order \cite{Papenbrock:1994kf,Arnone:2002yh,Arnone:2003pa,Codello:2013bra}. 
As with strongly interacting problems in statistical and quantum field theory \cite{Berges:2000ew}, the value of exact RG equations is that they provide the setting for simple approximations able to capture non-perturbative effects efficiently -- a real necessity in present day analysis of strongly coupled problems in GR. We conclude with a first application where we recompute the 1PN action in a very cheap and effective way: the correct result is obtained without the need for the gravitational three vertex -- thus avoiding the cumbersome part of the calculation by hedging on the inner nonlinearity of the RG equation.

\subsection*{\color{blue}An RG equation for GR}\vspace{-0.2cm}

Our goal is to develop a formalism based on the renormalization group (RG) equation that allows us to describe the conservative dynamics of two point particles interacting through gravity. We work in Lorentzian signature $\{- + + +\}$. With the gravitational coupling defined as \(\kappa = 32\pi G_N\), the system is governed by the action
\begin{equation}
S[g,x_1,x_2] = \frac{1}{\kappa}S_{\rm g}[g] + S_{\mathrm{\mathrm{pp}}}[g,x_1,x_2]\,,
\end{equation}
where $\frac{1}{\kappa}S_{\rm g}[g] = \frac{2c^4}{\kappa} \int dt \, d^3\mathbf{x} \,\sqrt{-g}\,R  + S_{\rm gf}$
is the Einstein–Hilbert action with a gauge fixing term added, and $S_{\mathrm{\mathrm{pp}}}[g,x_1,x_2] = - \sum_{n=1,2} m_n c^2 \int d\tau_n$ is the sum of the point-particle actions.  
The central object of interest is the \textit{effective action} \cite{Bernard:2015njp}, defined as the total action $S$ evaluated on the metric that solves the equations of motion:
\begin{equation*}
S_{\rm eff}[x_n] = S[g_*[x_n],x_n]\, ,
\end{equation*}
where $g_*$ satisfies
\begin{equation*}
\frac{\delta S}{\delta g_{\mu\nu}(x)}[g_*[x_n],x_n] = 0 \,.
\end{equation*}
In general, computing this effective action is highly non-trivial, and progress is typically made by expanding in powers of $\kappa$. This expansion is known as the Post-Minkowskian (PM) expansion.

Our aim is to construct a scale-dependent action $S_k[g,x_n]$ such that
\begin{equation*}
\lim_{k \to 0} S_k[\eta, x_n] = S_{\rm eff}[x_n]\,,
\end{equation*}
where $\eta$ denotes the Minkowski background. The action $S_k[g,x_n]$ should be obtained by solving an appropriate flow equation. The exact form of this equation can be heuristically derived in analogy with the Morris- Wetterich equation [?], starting from the one-loop correction to the effective action $\Gamma$. In our case, the first correction to the effective action is given by
\begin{equation*}
S_{\rm eff} = S_{\mathrm{\mathrm{pp}}} - \frac{\kappa}{2}\, S^{(1)}_{\mathrm{pp}} \cdot \, \Delta^{-1} \cdot \, S^{(1)}_{\mathrm{pp}} + \mathcal{O}(\kappa^2)\,,
\end{equation*}
evaluated at $g_{\mu\nu} = \eta_{\mu\nu}$ where $S^{(1)}_{\mathrm{pp}}=\frac{\delta S_{\mathrm{\mathrm{pp}}}}{\delta g}$. In the harmonic gauge, the propagator takes the form
\begin{equation*}
\Delta^{-1}_{\mu \nu \alpha \beta}(x,y)= c^{-3}\int \frac{d^4q}{(2\pi)^4} \frac{P_{\mu \nu \alpha \beta}}{-q^2} \, e^{-i  q \cdot(x-y) } \, ,
\end{equation*}
where $P_{\mu \nu \alpha \beta}=\frac{1}{2} \left(
\eta_{\mu\alpha} \eta_{\nu\beta} +
\eta_{\mu\beta} \eta_{\nu\alpha} -
\eta_{\alpha\beta} \eta_{\mu\nu}
\right)$. For brevity, summation over indices and integrals are represented by a ``$\cdot$''.

\begin{figure}
\centering
\includegraphics[height=16ex]{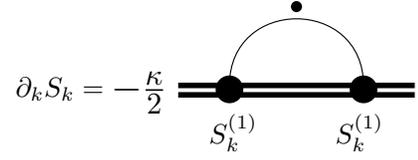}
\caption{Diagrammatic representation of the functional RG equation for GR. Each big dot over the double lines represent the first functional derivative of the running effective action $S_k$ for the two-particle system. The thin line corresponds to the regularized propagator $G_k$, and the dot above indicates the derivative $\partial_k$ acting on it.}
\label{RGforGR}
\end{figure}

Within the RG framework, the propagator is modified by introducing a $k$-dependent regulator function $R_k$. This leads to a scale-dependent effective action of the form
\begin{equation*}
S_{k}=S_{\mathrm{\mathrm{pp}}}-\frac{\kappa}{2}S^{(1)}_{\mathrm{pp}}  \cdot \Delta^{-1}_k \cdot S^{(1)}_{\mathrm{pp}}+\mathcal{O}(\kappa^2)\,,
\end{equation*}
where $\Delta_k=\Delta+R_k$. Differentiating this expression with respect to $k$, one obtains the approximate flow equation
\begin{equation*}
\partial_k S_{k}=-\frac{\kappa}{2}S^{(1)}_{\mathrm{pp}} \cdot  \partial_k \Delta^{-1}_k \cdot  S^{(1)}_{\mathrm{pp}}\,.
\end{equation*}
In the full functional renormalization group (FRG) framework, one can promote $S^{(1)}_{\mathrm{pp}}$ to the running first derivative of the effective action (RG improvement) and replace the Minkowski metric with an arbitrary metric $g_{\mu\nu}$. The flow equation then takes the form
\begin{equation}\label{eq:flowWett_old}
\boxed{\; 
\partial_k S_k[g] = - \frac{\kappa}{2} \, S_k^{(1)}[g] \cdot \partial_k G_k[g] \cdot S_k^{(1)}[g] \;} 
\end{equation}
which depends on the regularised propagator
\begin{equation*}
G_k[g] = (S^{(2)}_{\rm g}[g] + R_k)^{-1}\,.
\end{equation*}
The regulator takes the general form
\begin{equation*}
(R_k)^{\mu\nu,\alpha \beta} = c^4 R_k(-\bm{\partial}^2) P^{\mu\nu\alpha\beta}\,,
\end{equation*}
such that the momentum kernel depends on the three-momenta $\bm{p}^2$. The requirements on $R_k(\bm{p}^2)$ are
\begin{equation*}
\lim_{k \to 0} R_k(\bm{p}^2) = 0\, , \qquad\qquad \lim_{k \to \infty} R_k(\bm{p}^2) = -\infty \, .
\end{equation*}
Equation~\eqref{eq:flowWett_old} can be represented diagrammatically as in Figure \ref{RGforGR}.
This flow equation governs the evolution of the effective action, starting from the initial condition 
$\lim_{k \to \infty} S_k = S_{\mathrm{pp}}$ and integrating down to the limit $k \to 0$, where the full 
effective action is recovered upon setting $g_{\mu\nu} = \eta_{\mu\nu}$. 
In this way, the RG flow continuously interpolates between the free point-particle action and the effective dynamics of the interacting system. Note that there are no loops in the flow equation and no factors of $\hbar$.

\subsection*{\color{blue}Recovering the PM expansion}\vspace{-0.2cm}
The fundamental requirement of the flow equation is that it must exactly reproduce the standard perturbative series. To show this, we start by expanding the effective action at scale $k$ in powers of $\kappa$:
\begin{equation*}
S_k= S_{\mathrm{\mathrm{pp}}} + \kappa S_{1} + \kappa^{2} S_{2} + \kappa^{3} S_{3} + \dots 
\end{equation*}
Here, all PM corrections are implicitly $k$-dependent. We will also omit the explicit $k$-dependence on $G$ for brevity. Substituting this expansion into the flow equation \eqref{eq:flowWett_old}, we obtain flows at each order in $\kappa$. At first post-Minkowskian (1PM) order, the flow reduces to:
\begin{equation*}
\partial_{k}S_{1}= -\frac{1}{2}S_{\mathrm{pp}}^{(1)} \cdot  \partial_{k}G \cdot  S_{\mathrm{pp}}^{{(1)}}\,.
\end{equation*}
The dependence on $k$ appears only through a total derivative. Integrating both sides from $k=0$ to $k=\infty$ yields:
\begin{equation*}
    S_{1}(\infty)-S_{1}(0)=-\frac{1}{2}S_{\mathrm{pp}}^{(1)} \cdot \left[G_{\lowercase{AB}}(\infty)-G_{\lowercase{AB}}(0)\right] \cdot S_{\mathrm{pp}}^{(1)}\,.
\end{equation*}
Using the limiting behavior of $R_{k}$ and the initial condition $S_{k\to \infty }=S_{\mathrm{\mathrm{pp}}}$, we obtain the 1PM effective action:

\begin{equation*}
S_{1}(k=0)=-\frac{1}{2}S_{\mathrm{pp}}^{(1)} \cdot (\Delta^{-1}) \cdot S_{\mathrm{pp}}^{(1)}\,,
\end{equation*}
which coincides with the perturbative result. 
\noindent

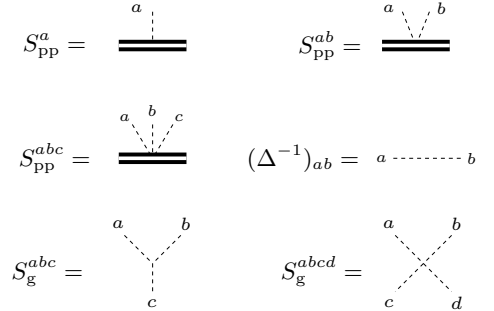
\begin{figure} [H]
\[
\scalebox{0.5}{
\begin{tikzpicture}[baseline=(current bounding box.center)]

  \node at (-6.5, 2) {\scalebox{2}{$\displaystyle S^{\lowercase{A}}_{\mathrm{pp}}=\ $}};
  \node at (-4.4, 3) {\scalebox{2.2}{$\displaystyle _a$}};
  \begin{scope}[shift={(-4,2)}, scale=1.8, transform shape]
    \begin{feynman}
      \vertex (i) at (-0.5,0);
      \vertex (d1) at (0,0.5);
       \vertex (d2) at (0,0);
      \vertex (f) at (0.5,0);
      \diagram* {     
       (d1) [dot]-- [scalar] (d2),
       (i) -- [double,line width=3pt, double distance=0.7ex] (f)
      };
    \end{feynman}
  \end{scope}

  \node at (0.82, 2) {\scalebox{2}{$\displaystyle S^{\lowercase{AB}}_{\mathrm{pp}}=\ $}};
  \node at (2.3, 3) {\scalebox{2.1}{$\displaystyle _a$}};
  \node at (3.7, 3) {\scalebox{2.1}{$\displaystyle _b$}};
  \begin{scope}[shift={(3,2)}, scale=1.8, transform shape]
    \begin{feynman}
      \vertex (i) at (-0.5,0);
      \vertex (d1) at (-0.2,0.5);
      \vertex (d3) at (0.2,0.5);
       \vertex (d2) at (0,0);
      \vertex (f) at (0.5,0);
      \diagram* {     
       (d1) [dot]-- [scalar] (d2),
       (d3) [dot]-- [scalar] (d2),
       (i) -- [double,line width=3pt, double distance=0.7ex] (f)
      };
    \end{feynman}
  \end{scope}

  \node at (-6.5, -1) {\scalebox{2}{$\displaystyle S^{\lowercase{ABC}}_{\mathrm{pp}}=\ $}};
  \begin{scope}[shift={(-4,-1)}, scale=1.8, transform shape]
    \begin{feynman}
    \node at (-0.4, 0.6) {\scalebox{1.1}{$\displaystyle _a$}};
    \node at (0, 0.7) {\scalebox{1.1}{$\displaystyle _b$}};
    \node at (0.4, 0.6) {\scalebox{1.1}{$\displaystyle _c$}};
      \vertex (i) at (-0.5,0);
      \vertex (d1) at (-0.3,0.5);
       \vertex (d0) at (0,0.5);
      \vertex (d3) at (0.3,0.5);
       \vertex (d2) at (0,0);
      \vertex (f) at (0.5,0);
      \diagram* {     
       (d1) [dot]-- [scalar] (d2),
       (d0) [dot]-- [scalar] (d2),
       (d3) [dot]-- [scalar] (d2),
       (i) -- [double,line width=3pt, double distance=0.7ex] (f)
      };
    \end{feynman}
  \end{scope}

\node at (-6.8, -4) {\scalebox{2}{$\displaystyle S_{\mathrm{g}}^{\lowercase{ABC}}=$}};
\begin{scope}[shift={(-4,-2.3)}, scale=1.5, transform shape]
  \begin{feynman}
  \node at (-0.6, -0.3) {\scalebox{1.5}{$\displaystyle _a$}};
  \node at (0.6, -0.3) {\scalebox{1.5}{$\displaystyle _b$}};
  \node at (0, -1.7) {\scalebox{1.5}{$\displaystyle _c$}};
    \vertex (i) at (-0.5,-0.5);
    \vertex (f) at (0.5,-0.5);
    \vertex (m) at (0,-1);
    \vertex (l) at (0,-1.5);
    \diagram* {     
      (i) -- [scalar] (m),
      (f) -- [scalar] (m),
      (m) -- [scalar] (l),
    };
  \end{feynman}
\end{scope}

  \node at (0, -1) {\scalebox{2}{$\displaystyle (\Delta^{-1})_{ab}=$}};
  \node at (2.1, -1) {\scalebox{2}{$\displaystyle _a$}};
  \node at (4.5, -1) {\scalebox{2}{$\displaystyle _b$}};
  \begin{scope}[shift={(3.3,-1)}, scale=1.8, transform shape]
    \begin{feynman}
      \vertex (i) at (-0.5,0);
      \vertex (f) at (0.5,0);
      \diagram* {     
       (i) -- [scalar] (f)
      };
    \end{feynman}
  \end{scope}

\node at (0.5, -4) {\scalebox{2}{$\displaystyle S_{\mathrm{g}}^{\lowercase{ABCD}}=$}};
\begin{scope}[shift={(3.2,-2.3)}, scale=1.5, transform shape]
  \begin{feynman}
  \node at (-0.6, -0.3) {\scalebox{1.5}{$\displaystyle _a$}};
  \node at (0.6, -0.3) {\scalebox{1.5}{$\displaystyle _b$}};
  \node at (-0.6, -1.7) {\scalebox{1.5}{$\displaystyle _c$}};
  \node at (0.6, -1.7) {\scalebox{1.5}{$\displaystyle _d$}};
    \vertex (i) at (-0.5,-0.5);
    \vertex (f) at (0.5,-0.5);
    \vertex (m) at (0,-1);
    \vertex (l) at (-0.5,-1.5);
     \vertex (k) at (0.5,-1.5);
    \diagram* {     
      (i) -- [scalar] (m),
      (f) -- [scalar] (m),
      (m) -- [scalar] (l),
      (m) -- [scalar] (k),
    };
  \end{feynman}
\end{scope}
\end{tikzpicture}
}
\]
\caption{Diagrammatic conventions used in the PM expansion. Thick double lines denote the worldlines of the sources and the single dashed lines denote graviton propagators}
\label{conventions}
\end{figure}

We now move to the second post-Minkowskian (2PM) order. Here, we make use of DeWitt indices such that lower case Latin indices include the spacetime indices and the coordinates e.g., $g_a = g_{\mu\nu}(x)$.
Then the $n$th functional derivative of a functional $F$ is denoted by $F^{a \dots a_n}$.
Repeated Latin indices imply a sum over spacetime indices and an integral over spacetime.
Then the relevant flow equation is
\begin{equation} \label{eq:2PM}
\partial_{k}S_{2}=-(S_{1}^{\lowercase{A}}S_{\mathrm{pp}}^{\lowercase{B}}) \, \partial_{k}G_{\lowercase{AB}}\,.
\end{equation}
In this case, both the propagator and the one-point function $S_{1}^{\lowercase{A}}$ depend on $k$. To determine $S_{1}^{\lowercase{A}}$, we differentiate the flow equation \eqref{eq:flowWett_old} with respect to the metric:
\begin{equation*}
\partial_{k}S_k^{a}= \frac{\delta }{\delta g_{a}}\partial_{k}S_k=-\frac{\kappa}{2}\frac{\delta }{\delta g_{a}}\left(S_k^{b}[g] \, \partial_{k}G_{bc}[g] \, S_k^{c}[g]\right)\,.
\end{equation*}
The functional derivative acts both on the one-point functions and on the inverse propagator, generating the graviton three-vertex $S_{\mathrm{g}}^{\lowercase{ABC}}$:
\begin{equation*}
\partial_{k}S_k^{\lowercase{A}} = -\kappa \,  S_k^{\lowercase{AB}}S_k^{\lowercase{C}} \, \partial_{k}G_{\lowercase{BC}} 
+ \frac{\kappa}{2}S_{\mathrm{g}}^{\lowercase{ABC}}S_k^{\lowercase{D}}S_k^{\lowercase{E}} \, \partial_{k}\!\left(G_{\lowercase{BD}}G_{\lowercase{CE}}\right)\,.
\end{equation*}
At 1PM order, integrating from $k$ to $\infty$ gives:
\begin{equation} \label{eq:S1A}
S^{a}_{1}=-S^{\lowercase{AB}}_{\mathrm{pp}}S^{\lowercase{C}}_{\mathrm{pp}}G_{\lowercase{BC}}+\frac{1}{2}S_{\mathrm{g}}^{\lowercase{ABC}}S^{\lowercase{D}}_{\mathrm{pp}}S^{\lowercase{E}}_{\mathrm{pp}}G_{\lowercase{BD}}G_{\lowercase{CE}}\,.
\end{equation}
Substituting this into \eqref{eq:2PM}, we find:
\begin{align*}
\partial_{k}S_{2}
&=\frac{1}{2}S^{\lowercase{AB}}_{\mathrm{pp}}S^{\lowercase{C}}_{\mathrm{pp}}S^{\lowercase{D}}_{\mathrm{pp}} \, \partial_{k}\!\left(G_{\lowercase{AC}}G_{\lowercase{BD}}\right)\\
&-\frac{1}{6}S_{\mathrm{g}}^{\lowercase{ABC}}S^{\lowercase{D}}_{\mathrm{pp}}S^{\lowercase{E}}_{\mathrm{pp}}S^{\lowercase{B}}_{\mathrm{pp}} \, \partial_{k}\!\left(G_{\lowercase{BD}}G_{\lowercase{CE}}G_{\lowercase{AB}}\right)\,.
\end{align*}
Thus, one obtains a total derivative that can be integrated straightforwardly, yielding the 2PM part of the effective action:
\begin{align} 
S_{2}(k=0)&=\tfrac{1}{2}S^{\lowercase{AB}}_{\mathrm{pp}}S^{\lowercase{C}}_{\mathrm{pp}}S^{\lowercase{D}}_{\mathrm{pp}}(\Delta^{-1})_{ac}(\Delta^{-1})_{bd} 
\nonumber \\[5pt]
&-\tfrac{1}{6}S_{\mathrm{g}}^{\lowercase{ABC}}S^{\lowercase{D}}_{\mathrm{pp}}S^{\lowercase{E}}_{\mathrm{pp}}S^{\lowercase{F}}_{\mathrm{pp}}(\Delta^{-1})_{ad}(\Delta^{-1})_{be}(\Delta^{-1})_{cf}\, . \nonumber
\end{align}
So we recover the known 2PM result, in agreement with previous literature. In the Supplementary Material, we show that the procedure also correctly reproduces the 3PM contribution $S_3$, thereby indicating that the iterative solution of the RG equation~\eqref{eq:flowWett_old} indeed reconstructs the full PM expansion. The complete result can be summarized diagrammatically as follows:
\begin{align*}
S_{\mathrm{eff}}&= S_{\mathrm{pp}} -\frac{\kappa}{2}  \,
\vcenter{\hbox{\raisebox{1ex}{\includegraphics[height=4ex]{unprop.tex}}}}
+\kappa^2 \ \left( \ \tfrac{1}{2}\,
\vcenter{\hbox{\includegraphics[height=4ex]{vertice2.tex}}}
-\tfrac{1}{6}\,
\vcenter{\hbox{\includegraphics[height=4ex]{vertice3.tex}}}
 \ \right)
 \\[2ex]
 & +\kappa^3 \ \left( \, 
\vcenter{\hbox{\includegraphics[height=20ex]{diag3PM.tex}}} \, \right)  \ +\mathcal{O}(\kappa^4) \, .
\end{align*}
These diagrams coincide exactly with the 3PM topologies reported in the literature \cite{Levi:2020pnEFT,Foffa:2013gsa} (conventions followed to draw the PM diagrams are depicted in Figure \ref{conventions}).

\subsection*{\color{blue}Fastrack to the 1PN Lagrangian}\vspace{-0.2cm}

In this section, we outline how the 1PN Lagrangian can be obtained directly from the flow equation~\eqref{eq:flowWett_old}. 
As is standard when solving any flow equation, we begin with a reasonable ansatz for $S_k$.  
The symmetries of the system, together with dimensional analysis, can be used to constrain the possible terms appearing in the final PN Lagrangian.  
We then promote each coefficient in this ansatz to a scale-dependent function of $k$.
The key to solving the flow equation is to identify $S^{(1)}_k$ with the scale-dependent stress-energy tensor:
\begin{equation*}
T^{\mu\nu}_k(x) = \frac{2}{\sqrt{-g}} \, \frac{\delta S_k}{\delta g_{\mu\nu}(x)} \, .
\end{equation*}
That is, all conserved quantities associated with the effective action (or Lagrangian) at scale $k$ must originate from this tensor.  
Evaluating equation \eqref{eq:flowWett_old} in the Minkowski background and expressing the propagator in Fourier space leads to:
{\small
\begin{equation} \label{eq:flow1PN}
\small\partial_{k} S_{k}
= \frac{\kappa}{8c^4} \int_{x,y} 
T_{k}^{\mu \nu}(x) \, P_{\mu \nu \alpha \beta} \, T_{k}^{\alpha \beta}(y)
\int_{q} \frac{e^{-i q \cdot (x-y)} \, \partial_{k} R_k(\mathbf{q}^2)}{\left(q^2 - R_k(\mathbf{q}^2)\right)^2} \, .
\end{equation}
}
\!\!\!\!Since $T^{00} = \mathcal{O}(c^2)$, $T^{0i} = \mathcal{O}(c)$, and $T^{ij} = \mathcal{O}(c^0)$, the contraction on the right-hand side becomes:
\begin{eqnarray*}
&&T^{\mu \nu}(x) P_{\mu \nu \alpha \beta} \, T^{\alpha \beta}(y)
= \frac12 T^{00}(x) T^{00}(y) - 2 \, T^{0i}(x) T^{0i}(y) \\
&&\qquad\qquad  + \frac12 T^{00}(x) T^{ii}(y) + \frac12 T^{00}(y) T^{ii}(x) 
+ \mathcal{O}(c^0) \, .
\end{eqnarray*}
To extract the 1PN effective action, we can neglect all terms of order $\mathcal{O}(c^0)$ in the above expression.  
The other relevant ingredient is the PN expansion of the momentum integral  $\int_{q} \frac{e^{-iq \cdot (x-y)} \, \partial_{k} R_k(\mathbf{q}^2)}
{\left(q^2 - R_k(\mathbf{q}^2)\right)^2} \, $,
which can be implemented systematically by expanding the denominator in powers of $q_0^2/c^2$:
\begin{equation*}
\frac{1}{\left(\mathbf{q}^2 -q_0^2/c^2 - R_k(\mathbf{q}^2)\right)^2}
\simeq \frac{ 1 + \frac{2q_0^2/c^2}{\mathbf{q}^2 - R_k(\mathbf{q}^2)} + \ldots }{\left(\mathbf{q}^2 - R_k(\mathbf{q}^2)\right)^2}\, .
\end{equation*}
This expansion is justified by the so-called \emph{method of regions} used in the EFT approach to classical gravity \cite{Rothstein:2003zf}. Note that the leading term in this series yields a Dirac delta $\delta(x_0 - y_0)$, corresponding to an instantaneous interaction.  
The subleading corrections in $q_0^2$ account for small departures from strict instantaneity.
All subsequent calculations are performed using the Litim regulator:$R_k(\mathbf{q}^2) = \left( \mathbf{q}^2-k^2 \right) \, \theta\!\left(k^2 - \mathbf{q}^2\right)
$ \cite{Litim:2000ci}.
The next step is to introduce the general 1PN ansatz (see Supplementary Material):
{\small
\begin{widetext}
\begin{align*}
S_{k} = \int {\rm d}t \, &\bigg\{
\frac{1}{2} m_1 \, \mathbf{v}_1^2 + \frac{1}{2} m_2 \, \mathbf{v}_2^2 
+ N_{k} \frac{G_N m_1 m_2}{R} 
+ \frac{1}{c^2} \bigg( 
\frac{m_1}{8} \mathbf{v}_1^4 + \frac{m_2}{8} \mathbf{v}_2^4 
+ \frac{G_N m_1 m_2}{R} \Big[ 
A_{k} (\mathbf{v}_1^2 + \mathbf{v}_2^2) 
+ B_{k} \, \mathbf{v}_1 \cdot \mathbf{v}_2 
\\
&
\!\!\!\!\!\!\!\!\!\!+ C_{k} (\mathbf{n} \cdot \mathbf{v}_1)(\mathbf{n} \cdot \mathbf{v}_2)
+ D_{k} \Big((\mathbf{n} \cdot \mathbf{v}_1)^2 + (\mathbf{n} \cdot \mathbf{v}_2)^2\Big) 
+ F_{k} \, (\mathbf{n} \cdot \mathbf{a}_1 - \mathbf{n} \cdot \mathbf{a}_2) \, R 
\Big] 
+ H_{k} \frac{G_N^2 m_1 m_2 (m_1 + m_2)}{R^2} \bigg) 
 \bigg\} + \mathcal{O}(\tfrac{1}{c^4}) \, ,
\end{align*}
\end{widetext}
}
\!\!\!\!\!\!where $R=|\mathbf{x}_1-\mathbf{x}_2|$ and $\mathbf{n}=\frac{\mathbf{x}_1-\mathbf{x}_2}{R}$. In this ansatz we introduce seven scale-dependent coefficients, which encode the running of the gravitational interaction terms. The corresponding leading-order equations of motion for this ansatz are $\mathbf{a}_1 = - N_k \, \frac{G m_2}{R^2} \, \mathbf{n} + \mathcal{O}(\tfrac{1}{c^2})$ and $\mathbf{a}_2 =  N_k \, \frac{G m_1}{R^2} \, \mathbf{n} + \mathcal{O}(\tfrac{1}{c^2}) $.
We can construct the relevant components of $T^{\mu \nu}$ by noting that all conserved quantities must be derivable from this tensor, and these can be obtained directly from the action $S_k$.  
Let us start with $T^{0i}$. For the 1PN computation we only require it to leading order, $\mathcal{O}(c)$. It can be determined from the total momentum:
\vspace{-0.27 em}
\begin{equation*}
\mathbf{P}^i = \frac{1}{c} \int {\rm d}^3\mathbf{x} \, T^{0i}(x)
= m_1 \mathbf{v}_1^i + m_2 \mathbf{v}_2^i + \mathcal{O}(\tfrac{1}{c^2}) \, .
\end{equation*}
Similarly, $T^{00}$ can be obtained up to $\mathcal{O}(c^0)$ from the total energy:
\begin{align*}
E &= \int {\rm d}^3\mathbf{x} \, T^{00}(x) 
= m_1 c^2 + m_2 c^2 \\
&\quad
+ \frac{1}{2} m_1 \mathbf{v}_1^2 + \frac{1}{2} m_2 \mathbf{v}_2^2  - N_k \frac{G_N m_1 m_2}{R} 
+ \mathcal{O}(\tfrac{1}{c^2}) \, .
\end{align*}
The case of $T^{ij}$ is different, since there is no direct conserved quantity associated with it.  
Instead, one can use the on-shell identity
\begin{equation*}
\int {\rm d}^3\mathbf{x} \, T^{ii}(x)
= \frac{1}{2 c^2} \frac{{\rm d}^2}{{\rm d}t^2} 
\int {\rm d}^3\mathbf{x} \, T^{00}(x) \, \mathbf{x}^2 \, ,
\end{equation*}
which follows from the conservation of the stress–energy tensor.  
This relates $T^{ii}$ to the $T^{00}$ previously found, once the equations of motion are used.
The resulting expressions are:
{\small
\begin{align*}
& T^{00}_k(\mathbf{x},t) =\sum_{n=1,2}\left[ m_nc^2+\tfrac{1}{2}m_nv_n^2-\tfrac{N_{k}}{2}\tfrac{G_N m_1m_2}{R}\right]\delta^{(3)}(\mathbf{x}-\mathbf{x}_n(t))
\\
&T^{0i}_k(\mathbf{x},t) =\sum_{n=1,2} m_n  c \ \mathbf{v}_n^i\,\delta^{(3)}(\mathbf{x}-\mathbf{x}_n(t))
\\
&T^{i j}_k(\mathbf{x},t)=\sum_{n=1,2}\left[m_n\mathbf{v}_n^i\mathbf{v}_n^j-\tfrac{N_k}{2}\tfrac{G_N m_1 m_2}{R^3}\mathbf{R}^i\mathbf{R}^j\right]
\delta^{(3)}(\mathbf{x}-\mathbf{x}_n(t))
\end{align*}
}
\!\!Inserting these expressions for $T^{\mu \nu}$ into equation \eqref{eq:flow1PN} and neglecting all self-energy divergencies (which can be consistently handled using dimensional regularization~\cite{Goldberger:2006bd}) yields:
\begin{widetext}
\begin{align} \label{eq:flowSk}
\partial_k S_{k} 
&= \frac{\kappa}{8} 
m_1 m_2 
\int {\rm d}t \int_{\mathbf{q}}
\left[ 1 + \frac{1}{c^2}\frac{2(\mathbf{v}_1 \cdot \mathbf{q})(\mathbf{v}_2 \cdot \mathbf{q})}{\mathbf{q}^2 - R_k(\mathbf{q}^2)} \right]
\frac{e^{-i\mathbf{q} \cdot \mathbf{R}} \, \partial_k R_k(\mathbf{q}^2)}
{\left(\mathbf{q}^2 - R_k(\mathbf{q}^2)\right)^2}  \\[4pt]
& + \frac{\kappa}{8c^2}m_1 m_2 \int {\rm d}t \,
\left[ \frac{3}{2} \left( \mathbf{v}_1^2 + \mathbf{v}_2^2 \right)
- 4  \, \mathbf{v}_1 \cdot \mathbf{v}_2
- N_k \frac{G_N (m_1 + m_2)}{R}  \right]
\int_{\mathbf{q}} \frac{e^{-i\mathbf{q} \cdot \mathbf{R} } \, \partial_k R_k(\mathbf{q}^2)}
{\left(\mathbf{q}^2 - R_k(\mathbf{q}^2)\right)^2}  
+ \mathcal{O}(\tfrac{1}{c^4})  \nonumber \, .
\end{align}
\end{widetext}
The coefficients associated to velocity and acceleration terms can be found directly using:
{\small
\begin{equation*}
\int_{0}^{\infty}\! {\rm d}k \int_{\mathbf{q}} 
  \frac{e^{-i \mathbf{q} \cdot \mathbf{R}}\, 
  \partial_k R_k(\mathbf{q}^2)}{(\mathbf{q}^2 - R_k(\mathbf{q}^2))^2}
= -\frac{1}{4\pi R}\,,
\end{equation*}
}
{\small
\begin{equation*}
\int_{0}^{\infty}\! {\rm d}k \int_{\mathbf{q}} 
\frac{e^{-i \mathbf{q} \cdot \mathbf{R}}\, 
\partial_k R_k(\mathbf{q}^2)\, \mathbf{q}^a \mathbf{q}^b}
{(\mathbf{q}^2 - R_k(\mathbf{q}^2))^3}
= - \frac{1}{16\pi R}
\left(\delta^{ab} - 
\frac{\mathbf{R}^a \mathbf{R}^b}{R^2}\right).
\end{equation*}
}

\noindent The only non–trivial running is that of $H_k$, which is linearly related to the flow of $N_k$.  
We can thus extract the flows of $N_k$ and $H_k$ by evaluating $\partial_k S_k$ at fixed particle positions:
\begin{align}
\small \left. \partial_{k}S_{k} \right|_{\substack{\mathbf{r_1} = \text{const.} \\ \mathbf{r_2} = \text{const.}}} 
&  \small = T \ \partial_kN_k\frac{G_N m_1m_2}{R} \notag
\\ 
& \small +T \ \partial_kH_k \ \frac{1}{c^2}\left(\frac{G_N^2m_1^2m_2}{R^2}+\frac{G_N^2m_2^2m_1}{R^2}\right) \notag \, .
\end{align}
By comparing the LHS to the RHS we arrive at the following flow equations:
\begin{equation*}
\begin{aligned}
&\partial_kN_k=- \sqrt{\frac{8}{\pi}} \ \frac{J_{3/2}(kR)}{k^{3/2}R^{1/2}}\,,
  \\[3ex]
&\partial_kH_k= \sqrt{\frac{8}{\pi}} \ \frac{J_{3/2}(kR)}{k^{3/2}R^{1/2}} \ N_k \, ,
\end{aligned}
\end{equation*}
where  $J_{\nu}(x)$ are the Bessel functions of the first kind. We first determine the running of $N_{k}$ and then use it to extract the running of $H_{k}$. The running of the other coefficients can be found by directly integrating equation \eqref{eq:flowSk}. Integrating the flows from the UV scale down to $k=0$ yields the following values for the 1PN coefficients:
\begin{equation*}
N_{k=0} = 1 \,, \quad
A_{k=0} = \tfrac{3}{2} \,, \quad
B_{k=0} = -\tfrac{7}{2} \,, \quad
C_{k=0} = -\tfrac{1}{2} \,,
\end{equation*}
\begin{equation*}
D_{k=0} = 0 \,, \quad
F_{k=0} = 0 \,, \quad
H_{k=0} = -\tfrac{1}{2} \,.
\end{equation*}
These values coincide exactly with the 1PN action in harmonic coordinates \cite{Blanchet:2002gz}.

\subsection*{\color{blue}Future directions}
\vspace{-0.2cm}

In this Letter we have proposed a new RG equation for GR and showed that it reproduces perturbation theory as well as being effective in practical PN calculations. In a companion paper we will prove that our equation is indeed exact beyond the perturbative domain as it is  directly related to the Polchinski RG \cite{Paper2}. Our equation can be contrasted with the approach of \cite{Morris:2016nda}, which generalises the Polchinski RG in order to preserve diffeomorphism invariance.  

After these formal developments, the key question that now arises is whether these RG equations are effective in computing and taking into account non-perturbative effects at the classical level. The direction that opens is clearly that of setting up a {\it functional} expansion for the effective action; for example, a derivative expansion or something similar adapted to the two-body problem, ultimately hoping that simple schemes can give profound analytical insights and even good estimates for physical observables.
We also signal possible applications -- among others -- to problems relevant to cosmology, as for example structure formation along the lines of \cite{Pietroni:2011iz,Floerchinger:2016hja}.

As a final note, we should comment that while in RG applications to quantum gravity \cite{Reuter:2019byg,Percacci:2017fkn} the main problem is always the lack of experimental observables, here we have clear gauge-invariant quantities that can be measured experimentally -- like the energy of the bound orbit or the total power irradiated. Thus, the effort of doing classical RG will also illuminate us on the `big brother' problem of quantum gravity by testing the general application of RG ideas  to the domain of gravity. It is also noteworthy that our equation is a Lorentzian RG for classical gravity, complementing the quantum counterparts \cite{Manrique:2011jc,Biemans:2016rvp,Fehre:2021eob,Banerjee:2022xvi,DAngelo:2022vsh,DAngelo:2023wje,Saueressig:2023tfy,Thiemann:2024vjx,Ferrero:2024rvi}.
\vspace{-0.2cm}

\subsection*{\color{blue}Acknowledgments}
\vspace{-0.2cm}


The authors acknowledge financial support from the CSIC grant I+D-2022-22520220100174UD. 
K.F. also acknowledges financial support from ANII-SNI-2022-1-1012554. 
F.G. acknowledges support from the Comisión Académica de Posgrados (CAP), Universidad de la República (Uruguay), through a Master's fellowship, 
and gratefully acknowledges the hospitality and support of the Ca’ Foscari University of Venice during a research internship.
\\


\appendix
\onecolumngrid
\section*{Supplementary Material}

\subsection{Details of the 3PM computation}

Now lets dive into the 3PM, for that one needs the equation
\beq \label{eq:3PM}
\partial_{k}S_{3}=-(S_{2}^{\lowercase{A}}S_{\mathrm{pp}}^{\lowercase{B}}+\frac{1}{2}S_{1}^{\lowercase{A}}S_{1}^{\lowercase{B}}) \ \partial_k G_{\lowercase{AB}}
\eeq
We already know $S^{a}_1$ from equation \eqref{eq:S1A}, so all we have left is to find $S_{2}^{\lowercase{A}}$. By taking a derivative of \eqref{eq:2PM} we have
\beq \label{eq:S2A}
\partial_{k}S^{a}_2=-S^{ab}_{1}S^{\lowercase{C}}_{\mathrm{pp}} \ \partial_k G_{\lowercase{BC}}-S^{\lowercase{AB}}_{\mathrm{pp}}S^{c}_{1} \ \partial_k G_{\lowercase{BC}}+S_{\mathrm{g}}^{\lowercase{ABC}}S^{\lowercase{D}}_{\mathrm{pp}}S^{e}_1 \ \partial_k\left(G_{\lowercase{BD}}G_{\lowercase{CE}}\right)
\eeq
Here we see that $S^{ab}_{1}$ is also needed for the computation, this can be found by differentiating again the flow equation \eqref{eq:flowWett_old} with respect to the metric. Then one gets
\begin{align*}
\partial_{k}S^{ab}=&-\kappa S^{abc}S^{d} \ \partial_k G_{\lowercase{CD}}-\kappa S^{ac}S^{bd} \ \partial_k G_{\lowercase{CD}}+\kappa S_{g}^{\lowercase{BCD}}S^{ae}S^{f} \ \partial_k\left(G_{\lowercase{CE}}G_{\lowercase{DF}}\right)\\
&+\frac{\kappa}{2}S_{\mathrm{g}}^{\lowercase{ABCD}}S^{e}S^{f} \ \partial_k\left(G_{\lowercase{DF}}G_{\lowercase{CE}}\right)+\kappa S_{\mathrm{g}}^{\lowercase{ACD}}S^{be}S^{f} \ \partial_k\left(G_{\lowercase{DF}}G_{\lowercase{CE}}\right)-\kappa S_{\mathrm{g}}^{\lowercase{ACD}}S^{e}S^{f}S_{\mathrm{g}}^{\lowercase{BHI}} \ \partial_k\left(G_{\lowercase{CH}}G_{\lowercase{DE}} G_{\lowercase{FI}}\right)
\end{align*}
If we take the 1PM part of the latter the only $k$ dependence is in the propagators so we are left with a total derivative which can be integrated, giving:
\begin{align} \label{eq:SAB1}
S^{\lowercase{AB}}_1=&- S^{\lowercase{ABC}}_{\mathrm{pp}}S^{\lowercase{D}}_{\mathrm{pp}} G_{\lowercase{CD}}-S_{\mathrm{pp}}^{ac}S_{\mathrm{pp}}^{bd}G_{\lowercase{CD}}+ S_{\mathrm{g}}^{\lowercase{BCD}}S_{\mathrm{pp}}^{ae}S_{\mathrm{pp}}^{f}G_{\lowercase{CE}}G_{\lowercase{DF}} \notag\\ 
&+\frac{1}{2}S_{\mathrm{g}}^{\lowercase{ABCD}}S_{\mathrm{pp}}^{e}S_{\mathrm{pp}}^{f}G_{\lowercase{DF}}G_{\lowercase{CE}}+ S_{\mathrm{g}}^{\lowercase{ACD}}S_{\mathrm{pp}}^{be}S_{\mathrm{pp}}^{f}G_{\lowercase{DF}}G_{\lowercase{CE}} \\
&- S_{\mathrm{g}}^{\lowercase{ACD}}S_{\mathrm{pp}}^{e}S_{\mathrm{pp}}^{f}S_{\mathrm{g}}^{\lowercase{BHI}}G_{\lowercase{CH}}G_{\lowercase{DE}} G_{\lowercase{FI}} \notag
\end{align}
Now one can introduce \eqref{eq:SAB1} and \eqref{eq:S1A} into \eqref{eq:S2A} to find $S^{a}_2$:
\begin{align} \label{eq:S2a}
S^{a}_2&=\frac{1}{2}S^{abc}_{\mathrm{g}}S^{\lowercase{D}}_{\mathrm{pp}} S^{efh}_{\mathrm{g}}S^{\lowercase{I}}_{\mathrm{pp}}S^{\lowercase{J}}_{\mathrm{pp}} G_{\lowercase{BD}}G_{\lowercase{CE}}G_{\lowercase{FI}}G_{\lowercase{HJ}}-\frac{1}{6}S^{abcd}_{\mathrm{g}}S^{\lowercase{E}}_{\mathrm{pp}} S^{\lowercase{F}}_{\mathrm{pp}}S^{\lowercase{H}}_{\mathrm{pp}}G_{\lowercase{BE}}G_{\lowercase{CF}}G_{\lowercase{DH}}\\
&-\frac{1}{2} S^{\lowercase{AB}}_{\mathrm{pp}}S^{\lowercase{C}}_{\mathrm{pp}}S_{\mathrm{g}}^{\lowercase{DEF}}S^{\lowercase{H}}_{\mathrm{pp}}G_{\lowercase{BD}}G_{\lowercase{EC}} G_{\lowercase{FH}}-S_{\mathrm{g}}^{\lowercase{ABC}}S^{\lowercase{D}}_{\mathrm{pp}}S^{\lowercase{EF}}_{\mathrm{pp}}S^{\lowercase{H}}_{\mathrm{pp}}G_{\lowercase{BD}}G_{\lowercase{CE}} G_{\lowercase{FH}}\notag\\
&+ S^{\lowercase{AB}}_{\mathrm{pp}} S^{\lowercase{CD}}_{\mathrm{pp}}S^{\lowercase{E}}_{\mathrm{pp}}G_{\lowercase{BC}}G_{\lowercase{DE}}+\frac{1}{2} S^{\lowercase{ABC}}_{\mathrm{pp}} S^{\lowercase{D}}_{\mathrm{pp}}S^{\lowercase{E}}_{\mathrm{pp}}G_{\lowercase{BD}}G_{\lowercase{CE}}\notag
\end{align}
Finally, we put \eqref{eq:S2a} and \eqref{eq:S1A} into \eqref{eq:3PM} to get the flow for $S_{3}$:
\[
\scalebox{0.65}{
\begin{tikzpicture}[baseline=(current bounding box.center)]
\node at (-10.4, 5) {\scalebox{1.5}{$\displaystyle\partial_{k}S_{3} \ =  - \ (S_{2}^{\lowercase{A}}S_{\mathrm{pp}}^{\lowercase{B}}+\frac{1}{2}S_{1}^{\lowercase{A}}S_{1}^{\lowercase{B}}) \  \ \partial_kG_{\lowercase{AB}} $}};
  \node at (-12.7, 2) {\scalebox{1.5}{$\displaystyle =- \ \frac{1}{8}$}};
  \begin{scope}[shift={(-11.2,2)}, scale=0.8, transform shape]
    \begin{feynman}
      \vertex (i) at (-0.5,0);
      \vertex (d1) at (0.5,0);
      \vertex (d2) at (1.5,0);
      \vertex (e1) at (1,1.5);
      \vertex (e2) at (3,1.5);
      \vertex (e22) at (2.5,0);
      \vertex (f1) at (3.5,0);
      \vertex (f) at (4.5,0);
      \diagram* {
        (d1) -- [plain] (e1) -- [plain] (d2),
        (e22) -- [plain] (e2) -- [plain] (f1),
        (e1) -- [plain] (e2),
        (i) -- [double, line width=3pt, double distance=0.5ex] (f)
      };
    \coordinate (mid) at  ($(e1)!0.5!(e2)$);
    \draw[line width=2pt]; 
   \fill (2,1.8) circle (3pt); 
    \end{feynman}
  \end{scope}

 \node at (-6.9, 2) {\scalebox{1.5}{$\displaystyle - \ \frac{1}{2}$}};
  \begin{scope}[shift={(-5.8,2)}, scale=0.8, transform shape]
    \begin{feynman}
      \vertex (i) at (-0.5,0);
      \vertex (d1) at (0.5,0);
      \vertex (d2) at (1.5,0);
      \vertex (e1) at (1,1.5);
      \vertex (e2) at (3,1.5);
      \vertex (e22) at (2.5,0);
      \vertex (f1) at (3.5,0);
      \vertex (f) at (4.5,0);
     \coordinate (mid) at ($(d1)!0.5!(e1)$);
      \diagram* {
        (d1) -- [plain] (e1) -- [plain] (d2),
        (e22) -- [plain] (e2) -- [plain] (f1),
        (e1) -- [plain] (e2),
        (i) -- [double, line width=3pt, double distance=0.5ex] (f)
      };
     \draw[line width=2pt]; 
   \fill (0.5,1) circle (3pt); 
    \end{feynman}
  \end{scope}

 \node at (-1.5, 2) {\scalebox{1.5}{$\displaystyle + \ \frac{1}{6}$}};
  \begin{scope}[shift={(1.2,2)}, scale=0.8, transform shape]
    \begin{feynman}
      \vertex (i) at (-2.5,0);
      \vertex (v) at (0,1.5);
      \vertex (a) at (-1.5,0);
      \vertex (b) at (-0.5,0);
      \vertex (c) at (0.5,0);
      \vertex (d) at (1.5,0);
      \vertex (f) at (2.5,0);
      \coordinate (mid) at ($(b)!0.5!(v)$);
      \diagram* {
        (v) -- [plain] (a),
        (v) -- [plain] (b),
        (v) -- [plain] (c),
        (v) -- [plain] (d),
        (i) -- [double, line width=3pt, double distance=0.5ex] (f),
      };
        \draw[line width=2pt]; 
          \fill (-0.9,1) circle (3pt);
    \end{feynman}
  \end{scope}

 \node at (-12, -1) {\scalebox{1.5}{$\displaystyle + \ $}};
  \begin{scope}[shift={(-9.5,-1)}, scale=0.8, transform shape]
    \begin{feynman}
      \vertex (i) at (-2.5,-0.15);
      \vertex (i1) at (-2,0);
      \vertex (i0) at (-0.7,0);
      \vertex (v1) at (0.6,0);
      \vertex (f1) at (2,0);
      \vertex (f) at (2.5,-0.15);
      \vertex (top) at (-0.7,1.5);
      \vertex (loopend) at (2,0);
      \coordinate (mid) at ($(i1)!0.5!(top)$);
      \diagram* {
        (i1) -- [plain] (v1) -- [plain] (f1),
        (v1) -- [plain] (top),
        (top) -- [plain] (i1),
        (top) -- [plain] (i0),
        (top) -- [plain] (v1),
        (v1) -- [plain, half left, looseness=1.3] (loopend),
        (i) -- [double, line width=3pt, double distance=0.5ex] (f),
      };
    \draw[line width=2pt]; 
    \fill (-1.5,1) circle (3pt);
    \end{feynman}
  \end{scope}

  \node at (-6.8, -1) {\scalebox{1.5}{$\displaystyle + \ \frac{1}{2}$}};
  \begin{scope}[shift={(-4.1,-1)}, scale=0.8, transform shape]
    \begin{feynman}
      \vertex (i) at (-2.5,-0.15);
      \vertex (i1) at (-2,0);
      \vertex (i0) at (-0.7,0);
      \vertex (v1) at (0.6,0);
      \vertex (f1) at (2,0);
      \vertex (f) at (2.5,-0.15);
      \vertex (top) at (-0.7,1.5);
      \vertex (loopend) at (2,0);
       \coordinate (mid) at ($(v1)!0.5!(top)$);
      \diagram* {
        (i1) -- [plain] (v1) -- [plain] (f1),
        (v1) -- [plain] (top),
        (top) -- [plain] (i1),
        (top) -- [plain] (i0),
        (top) -- [plain] (v1),
        (v1) -- [plain, half left, looseness=1.3] (loopend),
        (i) -- [double, line width=3pt, double distance=0.5ex] (f),
      };
      \draw[line width=2pt]; 
   \fill (0.1,1) circle (3pt);
    \end{feynman}
  \end{scope}

  \node at (-1.4, -1) {\scalebox{1.5}{$\displaystyle + \ \frac{1}{2}$}};
  \begin{scope}[shift={(1.3,-1)}, scale=0.8, transform shape]
    \begin{feynman}
     \vertex (i) at (-2.5,-0.15);
      \vertex (i1) at (-2,0);
      \vertex (i0) at (-0.7,0);
      \vertex (v1) at (0.6,0);
      \vertex (f1) at (2,0);
      \vertex (f) at (2.5,-0.15);
      \vertex (top) at (-0.7,1.5);
      \vertex (loopend) at (2,0);
      \diagram* {
        (i1) -- [plain] (v1) -- [plain] (f1),
        (v1) -- [plain] (top),
        (top) -- [plain] (i1),
        (top) -- [plain] (i0),
        (top) -- [plain] (v1),
        (v1) -- [plain, half left, looseness=1.3] (loopend),
        (i) -- [double, line width=3pt, double distance=0.5ex] (f),
      };
      \coordinate (mid) at (1.3,0.53);
    \draw[line width=2pt]; 
    \fill (1.3,0.8) circle (3pt);
    \end{feynman}
  \end{scope}

    \node at (-11.98, -4) {\scalebox{1.5}{$\displaystyle - \ $}};
  \begin{scope}[shift={(-10.6,-4)}, scale=0.8, transform shape]
    \begin{feynman}
     \vertex (i) at (-1,0);
    \vertex (v1) at (0,0.15);
    \vertex (v2) at (1,0.15);
    \vertex (v3) at (2,0.15);
    \vertex (v4) at (3,0.15);
    \vertex (f) at (4,0);

    \diagram* {
      (i) -- [double, line width=3pt, double distance=0.5ex] (f),

      (v1) -- [plain, half left, looseness=3] (v2),
      (v2) -- [plain, half left, looseness=3] (v3),
      (v3) -- [plain, half left, looseness=3] (v4),
    };
     \coordinate (mid) at (0.5,0.87);
    \draw[line width=2pt];
     \fill (0.5,1.3) circle (3pt);
    \end{feynman}
  \end{scope}

    \node at (-6.75, -4) {\scalebox{1.5}{$\displaystyle - \ \frac{1}{2}$}};
  \begin{scope}[shift={(-5.2,-4)}, scale=0.8, transform shape]
    \begin{feynman}
     \vertex (i) at (-1,0);
    \vertex (v1) at (0,0.15);
    \vertex (v2) at (1,0.15);
    \vertex (v3) at (2,0.15);
    \vertex (v4) at (3,0.15);
    \vertex (f) at (4,0);

    \diagram* {
      (i) -- [double, line width=3pt, double distance=0.5ex] (f),

      (v1) -- [plain, half left, looseness=3] (v2),
      (v2) -- [plain, half left, looseness=3] (v3),
      (v3) -- [plain, half left, looseness=3] (v4),
    };
     \coordinate (mid) at (1.5,0.87);
    \draw[line width=2pt];
    \fill (1.5,1.3) circle (3pt);
    \end{feynman}
  \end{scope}
  
  \node at (-1.3, -4) {\scalebox{1.5}{$\displaystyle - \ \frac{1}{2}$}};
  \begin{scope}[shift={(0.64,-4)}, scale=0.8, transform shape]
    \begin{feynman}
     \vertex (i0) at (-1.5,0);
      \vertex (i) at (-1,0);
    \vertex (v1) at (0,0);
    \vertex (v2) at (1.5,0);
    \vertex (v3) at (2,0);
    \vertex (f) at (3,0);
     \vertex (f0) at (3.5,0);

    \diagram* {

      (i) -- [plain, out=90, in=90, looseness=1.8] (v3),

      (v1) -- [plain, out=100, in=80, looseness=1] (v3),

      (v3) -- [plain, half left, looseness=2] (f),
      (i0) -- [double,line width=3pt, double distance=0.5ex] (f0),
    };
    \coordinate (mid) at (0.53,1.57);
    \draw[line width=2pt];
    \fill (0.5,1.9) circle (3pt);
    \end{feynman}
  \end{scope}

\end{tikzpicture}
}
\]
Remember that the dots over the lines represent the $\partial_k$ derivative acting on the modified propagators $G_{\lowercase{k}}$ (represented by a continuous line). By reducing all terms to a total derivative we can arrive at the final result for the 3PM effective action:
\[
\scalebox{0.7}{
\begin{tikzpicture}[baseline=(current bounding box.center)]

  \node at (-6.5, 2) {\scalebox{1.5}{$\displaystyle S_{3}=- \ \frac{1}{8}$}};
  \begin{scope}[shift={(-4.5,2)}, scale=0.8, transform shape]
    \begin{feynman}
      \vertex (i) at (-0.5,0);
      \vertex (d1) at (0.5,0);
      \vertex (d2) at (1.5,0);
      \vertex (e1) at (1,1.5);
      \vertex (e2) at (3,1.5);
      \vertex (e22) at (2.5,0);f
      \vertex (f1) at (3.5,0);
      \vertex (f) at (4.5,0);
      \diagram* {
        (d1) -- [scalar] (e1) -- [scalar] (d2),
        (e22) -- [scalar] (e2) -- [scalar] (f1),
        (e1) -- [scalar] (e2),
        (i) -- [double,line width=3pt, double distance=0.5ex] (f)
      };
    \end{feynman}
  \end{scope}

  \node at (-0.1, 2) {\scalebox{1.5}{$\displaystyle + \ \frac{1}{24}$}};
  \begin{scope}[shift={(2.8,2)}, scale=0.8, transform shape]
    \begin{feynman}
      \vertex (i) at (-2.5,0);
      \vertex (v) at (0,1.5);
      \vertex (a) at (-1.5,0);
      \vertex (b) at (-0.5,0);
      \vertex (c) at (0.5,0);
      \vertex (d) at (1.5,0);
      \vertex (f) at (2.5,0);
      \diagram* {
        (v) -- [scalar] (a),
        (v) -- [scalar] (b),
        (v) -- [scalar] (c),
        (v) -- [scalar] (d),
        (i) -- [double,line width=3pt, double distance=0.5ex] (f)
      };
    \end{feynman}
  \end{scope}

  \node at (-2.7, -4) {\scalebox{1.5}{$\displaystyle - \ \frac{1}{6}$}};
  \begin{scope}[shift={(-0.66,-4)}, scale=0.8, transform shape]
    \begin{feynman}
     \vertex (i0) at (-1.5,0);
      \vertex (i) at (-1,0);
    \vertex (v1) at (0,0);
    \vertex (v2) at (1.5,0);
    \vertex (v3) at (2,0);
    \vertex (f) at (3,0);
     \vertex (f0) at (3.5,0);

    \diagram* {
      (i) -- [scalar, out=90, in=90, looseness=1.8] (v3),

      (v1) -- [scalar, out=100, in=80, looseness=1] (v3),

      (v3) -- [scalar, half left, looseness=2] (f),
      (i0) -- [double,line width=3pt, double distance=0.5ex] (f0),
    };
    \end{feynman}
  \end{scope}

  \node at (-5.83, -1) {\scalebox{1.5}{$\displaystyle + \ \frac{1}{2}$}};
  \begin{scope}[shift={(-2.8,-1)}, scale=0.8, transform shape]
    \begin{feynman}
      \vertex (i) at (-2.5,-0.15);
      \vertex (i1) at (-2,0);
      \vertex (i0) at (-0.7,0);
      \vertex (v1) at (0.6,0);
      \vertex (f1) at (2,0);
      \vertex (f) at (2.5,-0.15);
      \vertex (top) at (-0.7,1.5);
      \vertex (loopend) at (2,0);
      \diagram* {
        (i1) -- [scalar] (v1) -- [scalar] (f1),
        (v1) -- [scalar] (top),
        (top) -- [scalar] (i1),
        (top) -- [scalar] (i0),
        (top) -- [scalar] (v1),
        (v1) -- [scalar, half left, looseness=1.3] (loopend),
        (i) -- [double,line width=3pt, double distance=0.5ex] (f),
      };
    \end{feynman}
  \end{scope}

  \node at (-0.1, -1) {\scalebox{1.5}{$\displaystyle - \ \frac{1}{2}$}};
  \begin{scope}[shift={(1.7,-1)}, scale=0.8, transform shape]
    \begin{feynman}
     \vertex (i) at (-1,-0.15);
    \vertex (v1) at (0,0);
    \vertex (v2) at (1,0);
    \vertex (v3) at (2,0);
    \vertex (v4) at (3,0);
    \vertex (f) at (4,-0.15);

    \diagram* {
      (i) -- [double,line width=3pt, double distance=0.5ex] (f),

      (v1) -- [scalar, half left, looseness=3] (v2),
      (v2) -- [scalar, half left, looseness=3] (v3),
      (v3) -- [scalar, half left, looseness=3] (v4),
    };
    \end{feynman}
  \end{scope}
\end{tikzpicture}
}
\]

\subsection{The 1PN Ansatz}

We now describe the procedure for constructing a complete Post-Newtonian (PN) ansatz for the conservative effective action of the two-body problem in General Relativity. We take the effective action to be a \emph{local functional} of the particle trajectories,
\begin{equation*}
S_{\mathrm{eff}} = F\!\left[ \mathbf{x}_1(t), \mathbf{x}_2(t), \dot{\mathbf{x}}_1(t), \dot{\mathbf{x}}_2(t), \ddot{\mathbf{x}}_1(t), \ddot{\mathbf{x}}_2(t), \ldots \right],
\end{equation*}
which may, in principle, depend on arbitrary time derivatives of the positions.

\subsubsection*{Symmetries of the Conservative Sector}

The effective action must respect the symmetries inherited from General Relativity, after gauge fixing to a PN-adapted coordinate system (e.g. harmonic gauge). These are:

\begin{itemize}
    \item \textbf{Spatial translations:}  
    $\mathbf{x} \rightarrow \mathbf{x} + \mathbf{a}$ \\
    This implies that the dependence on the positions can only appear through the relative separation  
    \[
    \mathbf{R} = \mathbf{x}_1 - \mathbf{x}_2 .
    \]
    
    \item \textbf{Spatial rotations:}  
    Only rotational scalars may appear in the action. The allowed building blocks include
    \[
    r^2,\quad \mathbf{v}_1^2,\quad \mathbf{v}_2^2,\quad \mathbf{v}_1 \cdot \mathbf{v}_2,\quad 
    \mathbf{n} \cdot \mathbf{v}_1,\quad \mathbf{n} \cdot \mathbf{v}_2, \, \ldots
    \]
    
    \item \textbf{Exchange symmetry $1 \leftrightarrow 2$:}  
    Interchanging particle labels must leave the action invariant, which imposes equality of coefficients in symmetric terms.  
    For example, if the action contains
    \[
    S_{\mathrm{eff}} \supset \int dt \left[ 
    \alpha \frac{G_N^2 m_1^2 m_2}{c^2 R^2} 
    + \beta \frac{G_N^2 m_2^2 m_1}{c^2 R^2} \right],
    \]
    then exchange symmetry requires $\alpha = \beta$.
    
    \item \textbf{Time-reversal invariance:}  
    $t \rightarrow -t,\ \mathbf{v}_a \rightarrow -\mathbf{v}_a$ \\
    The conservative dynamics is invariant under time reversal, which forbids odd powers of velocities or accelerations. For instance, a term like
    \[
    \frac{G_N m_1 m_2\, \mathbf{n} \cdot \mathbf{v}_1}{c\, R}
    \]
    is excluded. This implies that the conservative action contains only even powers of $1/c$.
    
    \item \textbf{Boost invariance:}  
    The conservative dynamics must also be invariant under infinitesimal Lorentz boosts. In the PN-expanded theory, this appears as a constraint among coefficients of different velocity structures, ensuring that the change of the action under a boost is a total time derivative.
\end{itemize}

\subsubsection*{General Structure}

A generic building block without higher derivatives can be written as
\begin{equation*}
G_N^{\alpha} m_1^{\beta} m_2^{\gamma} R^{\rho} 
(\mathbf{v}_1^2)^{\sigma} (\mathbf{v}_2^2)^{\delta} 
(\mathbf{n} \cdot \mathbf{v}_1)^{\epsilon} (\mathbf{n} \cdot \mathbf{v}_2)^{\nu} ,
\end{equation*}
where all exponents are positive integers except for $\rho$, which can be a positive or negative integer to account for the singular behavior as $R \rightarrow 0$.

Dimensional analysis, together with the above symmetries, constrains the allowed exponents and coefficients. For instance, the most general local ansatz for the 1PN conservative action takes the form

\begin{align*}
S_{\mathrm{eff}} &= \int dt \bigg[
\frac{1}{2} m_1 \mathbf{v}_1^2 + \frac{1}{2} m_2 \mathbf{v}_2^2 
+ N \frac{G_N m_1 m_2}{R} 
+ \frac{1}{c^2} \bigg( 
\frac{m_1}{8} \mathbf{v}_1^4 + \frac{m_2}{8} \mathbf{v}_2^4 
+ \frac{G_N m_1 m_2}{R} \big[ 
A (\mathbf{v}_1^2 + \mathbf{v}_2^2) 
+ B \, \mathbf{v}_1 \cdot \mathbf{v}_2 
+ C (\mathbf{n} \cdot \mathbf{v}_1)(\mathbf{n} \cdot \mathbf{v}_2) \\
&\qquad\qquad + D \big((\mathbf{n} \cdot \mathbf{v}_1)^2 + (\mathbf{n} \cdot \mathbf{v}_2)^2\big) 
+ F \, (\mathbf{n} \cdot \mathbf{a}_1 - \mathbf{n} \cdot \mathbf{a}_2) \, R 
\big]
+ H \left( \frac{G_N^2 m_1^2 m_2}{R^2} + \frac{G_N^2 m_2^2 m_1}{R^2} \right) 
\bigg) 
 \bigg] + \mathcal{O}(\tfrac{1}{c^4}).
\end{align*}

\end{document}